\documentclass[showpacs,aps,prx,floatfix,reprint,superscriptaddress,footinbib]{revtex4-2}
\usepackage{amssymb,amsfonts,amsmath}
\usepackage{graphicx}
\usepackage{verbatim}
\usepackage[utf8]{inputenc}
\usepackage[T1]{fontenc}
\usepackage{soul,xcolor,colortbl}
\usepackage{multirow}
\usepackage{bm}
\usepackage{array}
\usepackage{color}
\usepackage{hyperref} \hypersetup{colorlinks=true,citecolor=blue,linkcolor=blue,urlcolor=blue}
\usepackage{mdframed}
\usepackage{lineno}
\usepackage{setspace}
\usepackage{makecell}

\definecolor{pranab_green}{rgb}{0.31,0.53,0.10}
\definecolor{pranab_red}{rgb}{0.85,0.23,0.11}

\newcolumntype{L}[1]{>{\raggedright\arraybackslash}p{#1} }
\newcolumntype{C}[1]{>{\centering  \arraybackslash}p{#1} }
\newcolumntype{R}[1]{>{\raggedleft \arraybackslash}p{#1} }

\def\AFLOW{{\small AFLOW}}

\def\RMSrD{{\small RMSrD}}

\def\AFLOW{{\small AFLOW}}

\makeatletter \renewcommand\frontmatter@abstractwidth{\dimexpr\textwidth\relax} \makeatother

\begin{document}
\title{Design rules for the thermal and elastic properties of rare-earth disilicates}

\author{Cormac Toher}
\email[]{cormac.toher@utdallas.edu}
\affiliation{Dept. Materials Science and Engineering and Dept. Chemistry and Biochemistry, University of Texas at Dallas, Richardson, TX 75080, USA}
\affiliation{Center for Autonomous Materials Design, Duke University, Durham, NC 27708, USA}
\author{Mackenzie J. Ridley}
\affiliation{Dept. Materials Science and Engineering, University of Virginia, Charlottesville, VA 22904, USA}
\affiliation{Materials Science and Technology Division, Oak Ridge National Laboratory, Oak Ridge, TN 37831, USA}
\author{Kathleen Q. Tomko}
\affiliation{Dept. Materials Science and Engineering, University of Virginia, Charlottesville, VA 22904, USA}
\author{David Hans Olson}
\affiliation{Laser Thermal, Inc., Charlottesville, VA 22904, USA}
\author{Stefano Curtarolo}
\affiliation{Dept. Mechanical Engineering and Materials Science, Duke University, Durham, NC 27708, USA}
\affiliation{Center for Autonomous Materials Design, Duke University, Durham, NC 27708, USA}
\author{Patrick E. Hopkins}
\affiliation{Dept. Materials Science and Engineering, University of Virginia, Charlottesville, VA 22904, USA}
\affiliation{Dept. Mechanical and Aerospace Engineering, University of Virginia, Charlottesville, VA 22904, USA}
\affiliation{Dept. Physics, University of Virginia, Charlottesville, VA 22904, USA}
\author{Elizabeth J. Opila}
\affiliation{Dept. Materials Science and Engineering, University of Virginia, Charlottesville, VA 22904, USA}
\affiliation{Dept. Mechanical and Aerospace Engineering, University of Virginia, Charlottesville, VA 22904, USA}

\date{\today}

\begin{abstract}
\noindent
Rare-earth silicates are the current standard material for use as environmental barrier coatings for SiC-based ceramic matrix composites as hot-section components in gas-turbine engines.
Expanding the design space to all available rare-earth elements to facilitate optimizing functionality requires an understanding of systematic trends in $RE_2$Si$_2$O$_7$ properties.
In this work, we combine first-principles calculations with experimental measurements of Young's modulus, coefficient of thermal expansion, and thermal conductivity for a range of different $RE_2$Si$_2$O$_7$ compositions and phases.
Clear trends are observed in these properties as a function of the radius of the rare-earth cation.
In the case of Young's modulus and thermal expansion, these trends also hold for multi-component systems; while the thermal conductivity of multi-component systems is noticeably lower, indicating the potential of such materials to also act as thermal barriers.
These results provide design rules for developing new thermal and environmental barrier coatings with stiffness and thermal expansion engineered to match that of the substrate, while simultaneously having reduced thermal conductivity.

\end{abstract}
\maketitle

\section*{Introduction}

\noindent
Increased efficiency in gas turbine engines, used in aviation and power generation applications,
requires components to operate at higher temperatures where the Ni-based superalloys used for turbine blades and static components are no longer able to resist creep and deformation \cite{Perepezko_HotEngine_Science_2009, Padture_Science_2002}.
Therefore, new materials based on silicon carbide fiber reinforced SiC ceramic matrix composites (CMCs) are being developed for use in gas turbine engine hot-section components \cite{Jacobson_SiCeramics_JAmCeramSoc_1993, Spriet_CMC_KeyEngMater_1997, Beesley_CMC_KeyEngMater_1997, Naslain_CMC_MRSBull_2003, Zok_CMCTurbine_AmCeramSocBull_2016, Steibel_CMCTurbine_AmCeramSocBull_2019}.
However, SiC is vulnerable to corrosion by hot gases in the combustion chamber: O$_2$ reacts with SiC to produce SiO$_2$, which then reacts with H$_2$O combustion products to form Si(OH)$_4$ \cite{Jacobson_SiCeramics_JAmCeramSoc_1993, Opila_SiC_JAmCeramSoc_1997, Smialek_SiCSiN_AdvCompMater_1999}.
Environmental barrier coatings are hence required to protect SiC components from corrosion \cite{Lee_EBC_SurfCoatTech_2000, Xu_EBC_CeramInt_2017}.

In addition to being corrosion resistant, environmental barrier materials need to be thermodynamically stable, and have thermal and mechanical properties commensurate with the underlying substrate \cite{Turcer_ScrMater_Silicate_2018}.
It is particularly important that the elastic moduli and coefficient of thermal expansion (CTE) of any coating match that of the substrate, to prevent expansion mismatches that could cause cracking and delamination \cite{Turcer_ScrMater_Silicate_2018}.
Low thermal conductivity is also advantageous --- enabling materials to be used as both thermal and environmental barrier coatings \cite{Padture_Science_2002}.

Rare-earth disilicates ($RE_2$Si$_2$O$_7$) represent the current standard for environmental barrier coating (EBC) materials \cite{Lee_EBC_SurfCoatTech_2000, Luo_Silicates_JEurCeramSoc_2018, Ren_SCrMat_2019, Dong_JECerS_2019} for use with SiC CMCs \cite{Padture_Science_2002}.
The $\beta$ phase $RE_2$Si$_2$O$_7$ do not display polymorph transitions in the temperature ranges of interest to EBC materials \cite{Felsche_Silicate_JLessCommMet_1970, Felsche_Silicates_1973}, and additionally have thermal expansion coefficients that match that of SiC \cite{Turcer_ScrMater_Silicate_2018}.
Yet, only Sc$_2$Si$_2$O$_7$, Lu$_2$Si$_2$O$_7$, Yb$_2$Si$_2$O$_7$ and Tm$_2$Si$_2$O$_7$ are naturally stable in the $\beta$ phase \cite{Felsche_Silicate_JLessCommMet_1970,Felsche_Silicates_1973} for the temperature range of interest to turbine applications (up to $1704^\circ$C \cite{Dennis_NETL_DOE_Turbines_2016}).
Expanding EBC design space to all available rare-earth ($RE$) elements requires an understanding of systematic trends in $RE_2$Si$_2$O$_7$ properties.

High-entropy compositions can stabilize phases that would otherwise not be expected to exist \cite{curtarolo:art99, curtarolo:art140, Oses_NatureReview_2020, George_NRM_2019, hea1, gao2015high}.
Configurational disorder also suppresses thermal conductivity without compromising mechanical stiffness \cite{Braun_ESO_AdvMat_2018} --- useful for thermal protection barriers \cite{Ren_SCrMat_2019, Dong_JECerS_2019}.
For disilicates, undesirable high-temperature polymorphs have been avoided through phase stabilization of the low-CTE $\beta$ phase in both multi-component and high entropy $RE_2$Si$_2$O$_7$ solutions \cite{Fernandez_JSSChem_Silicates_2011, Dong_JECerS_2019, Sun_CorSci_HESilicates_2020}.
The properties of multi-component phases often follow a rule of mixtures, so knowledge of the thermal and mechanical properties of the components can be used to formulate design rules for high-entropy materials \cite{curtarolo:art140, Ridley_ActaMater_Silicates_2020, Oses_NatureReview_2020, George_NRM_2019}.
However, the thermomechanical and thermochemical properties of all $RE_2$Si$_2$O$_7$ for all 16 useable rare-earth elements (excluding radioactive Pm, including Sc and Y) are not fully known.
Therefore, design rules based on easily obtained or estimated quantities, such as ionic radius, are useful for predicting the properties of multi-component materials when some of the  components are unstable in the corresponding structure.
In this work, we combine density functional theory (DFT) with experimental methods to investigate trends in properties for rare-earth disilicates across the multiple available polymorphs.
The DFT calculations enable investigations of the thermo-mechanical properties of phases that are thermodynamically unstable as single-components, but which could form part of a high-entropy material.
The presented trends will inform materials design for next-generation multi-component rare-earth disilicate thermal and environmental barrier coatings.

\section*{Methods}

\subsection*{Computational procedures}

\noindent
Elastic and thermal properties were calculated using the Automatic
Elasticity Library (AEL \cite{curtarolo:art115}) and Automatic GIBBS
Library (AGL \cite{curtarolo:art96, BlancoGIBBS2004}) modules
implemented within the Automatic FLOW (AFLOW \cite{curtarolo:art65,
  aflow_fleet_chapter_full, curtarolo:art191, curtarolo:art190}) framework for computational materials design.
In AEL, 1\% and 0.5\% compressive and tensile normal and shear strains are applied to the structure in each independent direction, and the stress tensor for each strained structure is calculated using density functional theory (DFT) with the Perdew-Burke-Enzerhof (PBE) exchange-correlation functional \cite{PBE} as implemented in VASP \cite{vasp_prb1996}.
While other exchange-correlation functionals such as HSE \cite{Heyd2003} or SCAN \cite{Perdew_SCAN_PRL_2015} can give more accurate results, PBE has the advantage of being compatible with existing data sets in repositories such as \AFLOW\ \cite{curtarolo:art190, curtarolo:art104}, enabling the identification of more general trends in materials properties.
The stress-strain data are fitted to extract the elastic stiffness tensor, and the elastic constants are used to calculate the bulk and shear moduli in the Voigt and Reuss approximations and the Voigt-Reuss-Hill (VRH) average \cite{Hill_elastic_average_1952}.
The Young's modulus $E$ is calculated from the VRH bulk and shear moduli ($B_{\mathrm{VRH}}$ and $G_{\mathrm{VRH}}$) as $E = 9 B_{\mathrm{VRH}} G_{\mathrm{VRH}} / (3 B_{\mathrm{VRH}} + G_{\mathrm{VRH}})$.

In AGL, the energies for a set of isotropically compressed and expanded structures are calculated with VASP, and the energy-volume data is fitted to obtain the bulk modulus as a function of volume, and thus the Debye temperature and Gr{\"u}neisen parameter.
The Debye-Gr{\"u}neisen model is then used to calculate the vibrational contribution to the free energy, and this is combined with the DFT energies and the pressure-volume term to obtain the Gibbs free energy as a function of pressure, temperature and volume.
For each pressure-temperature point, the equilibrium volume is calculated by minimizing the Gibbs free energy.
Other thermodynamic quantities such as the heat capacity and thermal expansion are then calculated using the bulk modulus, Debye temperature and Gr{\"u}neisen parameter for the equilibrium volumes.
Lattice thermal conductivity is estimated using the Liebfried-Schl{\"o}mann model \cite{Leibfried_formula_1954, slack, Morelli_Slack_2006} as implemented within AGL \cite{curtarolo:art96, curtarolo:art115}: it should be noted that this model is generally only valid for materials with cubic symmetry and tends to underestimate thermal conductivities, but has been shown to be useful for estimating general trends \cite{curtarolo:art96, curtarolo:art115}.

DFT calculations are carried out using the AFLOW standard parameters \cite{curtarolo:art104}, with 8,000 to 10,000 {\bf k}-points per reciprocal atom, a basis set cut-off at least 1.4 times that recommended in the potential files, and the projector-augmented wave (PAW) potentials as listed in Ref. \cite{curtarolo:art104}.
Calculations are performed using both PBE and PBE+$U$: while the PBE+$U$ calculations for some systems are already available in the AFLOW database and the results often give better agreement with experiment, reliable $U$ values are not always readily available for all of the elements used here.
Therefore, to enable a comparison at the same level of theory so that trends can be observed, elasticity calculations are also performed without $U$ corrections to the extent possible.

\subsection*{Experimental procedures}

\noindent
High-purity pre-reacted $RE_2$Si$_2$O$_7$ powders (Praxair: Danbury, CT) were used to construct single and multi-component disilicate samples.
Powders were first heated in air at 900$^\circ$C to burn off organic impurities, followed by ball milling for 24 hours to promote powder mixing.
Powders were loaded into a 20 mm diameter graphite die and sintered via spark plasma sintering (SPS, DCS 25-10, Thermal Technologies, Santa Rosa, CA) in argon at 65 MPa and a maximum temperature between 1515 – 1620$^\circ$C.
Samples were then heat treated in a box furnace (CM Rapid Furnace, Bloomsfield NJ) in lab air at 1500$^\circ$C for 24 hours to restore oxygen stoichiometry.

Sample surfaces were polished with sub-micron polishing media prior to characterization and experimentation.
Structural phases were characterized using X-ray diffraction (Malvern Panalytical, Westborough, MA).
Material coefficients of thermal expansion (CTEs) were measured via dilatometry (Netzsch dil402c, Burlington, MA) up to 1250$^\circ$C in flowing argon.
Young's modulus was measured via nanoindentation with an MTS Nano XP (MTS Systems Corporation, Eden Prairie, MN) at room temperature.
Nanoindentations were taken 100 $\mu$m apart with 20-60 indents per sample at a constant strain rate of 0.1 s$^{-1}$.
A Poisson's ratio of 0.3 was assumed for all materials, which was in general agreement with the literature \cite{Turcer_ScrMater_Silicate_2018}.
Young's modulus was averaged for each sample indentation with data from 100-200 nm indentation depths to minimize interaction with any underlying porosity in the material.

Thermal conductivity was measured via steady-state thermo-reflectance (SSTR) laser technique \cite{Braun_RevSciInst_Thermoreflectance_2019}, where a split-beam laser acts as both a steady state heat source for the sample and as a subsequent probing laser to discern changes in the sample reflectance.
The thermoreflectance signal was then correlated to intrinsic thermal conductivity. All experimental data will be presented alongside computational results to discuss general trends in properties across the rare-earth series for $RE_2$Si$_2$O$_7$.

\subsection*{Rare-earth ionic radii}

\noindent
Rare-earth ($RE$) ionic radii data were compiled from Shannon database for each respective cation charge and coordination number \cite{Shannon_Radii_Database, Shannon_ActaCryst_Radii_1976}.
The coordination numbers of the $RE$ cation for different structure types are listed in Table \ref{tab:coordnum}.
In the case of phases where $RE$ cations on different sites have different coordinations, the ionic radii are calculated as the weighted average of the radii for each respective $RE$ cation site.

\begin{table*}[t!]
  \caption{\small Structural information including space group and coordination numbers for different $RE_2$Si$_2$O$_7$ structure types used to determine rare-earth ionic radii.
  }
  \label{tab:coordnum}
  {\footnotesize
    \begin{tabular}{|l|c|c|c|c|c|c|c|c|c|c|c|c|c|c|}
      \hline
     Structure type & Felsche reference \cite{Felsche_Silicate_JLessCommMet_1970} & Space group & AFLOW prototype label & Coordination number \\ \hline
                            &               A            &         76            &                                                          &         8                        \\
      $\alpha$        &               B            &           2         & A2B7C2\_aP44\_2\_4i\_14i\_4i       &              7.5              \\
       $\beta$         &              C            &         12         & A7B2C2\_mC22\_12\_aij\_h\_i       &               6                 \\
      $\gamma$     &              D            &         14         &                                                          &               6                \\
      $\delta$         &              E            &          62        & A7B2C2\_oP44\_62\_3c2d\_2c\_d  &               7                \\
               &              G            &     11    &            &               8                \\
      \hline
    \end{tabular}
  }
\end{table*}

\subsection*{Data analysis}

\noindent
The relationship between the $RE$ ionic radii and thermoelastic properties was quantified using the Pearson (linear) correlation.
The Pearson correlation $r$ between two variables, $X$ and $Y$, is calculated as:
\begin{equation}
  \label{Pearson}
  r = \frac{\sum_{i=1}^{n} \left(X_i - \overline{X} \right) \left(Y_i - \overline{Y} \right) }{ \sqrt{\sum_{i=1}^{n} \left(X_i - \overline{X} \right)^2} \sqrt{\sum_{i=1}^{n} \left(Y_i - \overline{Y} \right)^2}},
\end{equation}
where $\overline{X}$ and $\overline{Y}$ are the mean values of $X$ and $Y$.
Large magnitudes (close to 1) indicate a strong correlation, whereas values close to zero indicate a weak correlation.
A negative correlation indicates that one variable increases when the other decreases.

The discrepancy between quantities that should have the same value (e.g. rule of mixtures compared to measured values)
are also quantified using the normalized root-mean-square relative deviation (\RMSrD):
\begin{equation}
  \label{RMSD}
  {\small \mathrm{RMSrD}}  = \sqrt{\frac{ \sum_{i=1}^{n} \left( \frac{X_i - Y_i}{X_i} \right)^2 }{N - 1}}.
\end{equation}
Lower values of the \RMSrD\ indicate better agreement between the data sets being compared.

\section*{Results}

\subsection*{Structural characterization}

\begin{figure}[t!]
  \includegraphics[width=0.49\textwidth]{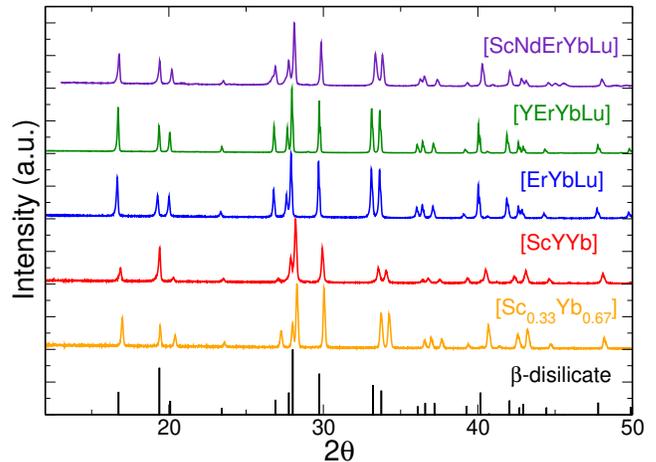}
  \vspace{-4mm}
  \caption{\small X-ray diffraction spectra for the multi-component rare-earth disilicates investigated in this work.
Secondary phases are present for (ScNdErYbLu)$_2$Si$_2$O$_7$, while the other compositions form homogeneous $\beta$-phases.The reference disilicate $\beta$-phase is for Yb$_2$Si$_2$O$_7$; peak shifting relative to this reference is related to changes in the rare earth cationic radii.}
  \label{fig:xrdspectra}
\end{figure}

\noindent
X-ray diffraction spectra for the multi-component disilicates investigated in this work are shown in Figure \ref{fig:xrdspectra}.
The compositions (Sc$_{0.33}$Yb$_{0.67}$)$_2$Si$_2$O$_7$, (ScYYb)$_2$Si$_2$O$_7$, (ErYbLu)$_2$Si$_2$O$_7$ and  (YErYbLu)$_2$Si$_2$O$_7$ were all found to form in the $\beta$-phase, with no secondary phases observed.
(ScNdErYbLu)$_2$Si$_2$O$_7$ was observed to form a secondary phase \cite{Ridley_Silicates_Mater_2022}.
The disilicate $\beta$-phase shown for reference in Figure \ref{fig:xrdspectra} is for Yb$_2$Si$_2$O$_7$; peak shifting relative to this reference is related to changes in the rare earth cation radii.

\subsection*{Elastic modulus}

\begin{figure*}[t!]
  \includegraphics[width=0.98\textwidth]{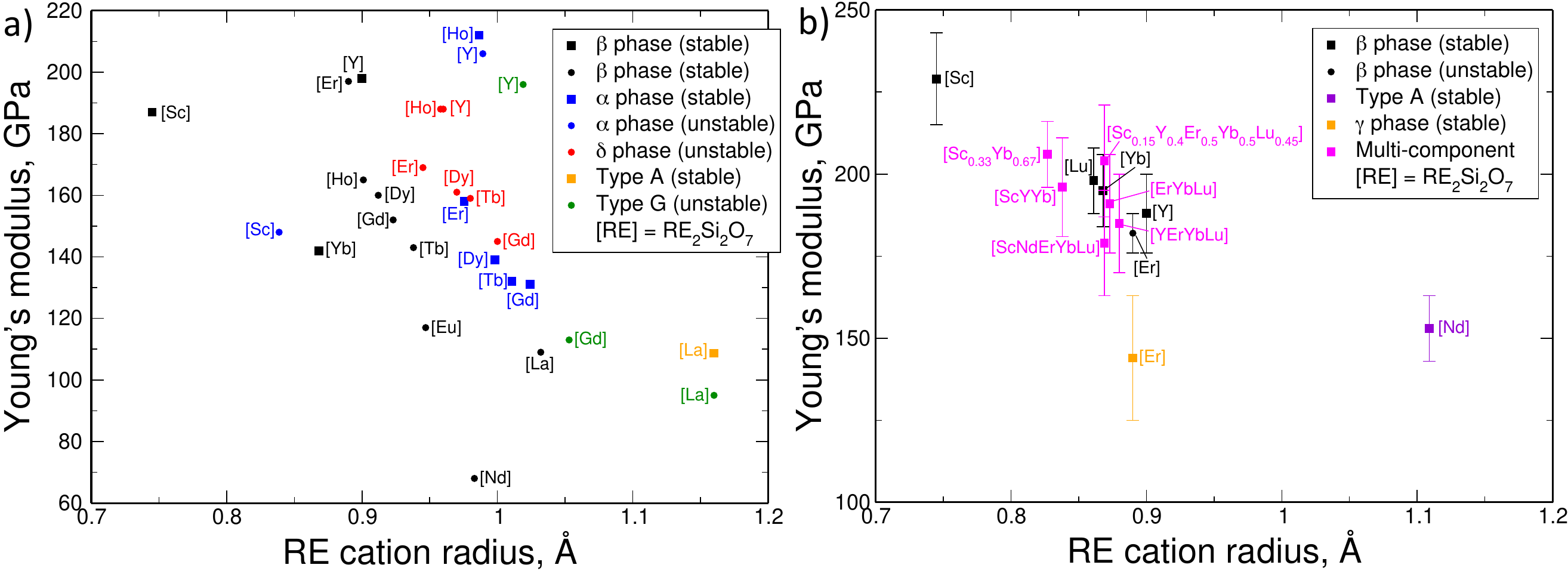}
  \vspace{-4mm}
  \caption{\small Young's modulus as a function of rare-earth ($RE$) cation radius for rare-earth disilicates obtained from
    {\bf (a)}  DFT calculations and
    {\bf (b)} experimental measurements.}
  \label{fig:elasticmod}
\end{figure*}

\noindent
Results for elastic moduli obtained from both DFT calculations and nanoindentation measurements as a function of $RE$ cation radius are shown in Figure \ref{fig:elasticmod}.
Values for single-component systems are listed in Table \ref{tab:singlecomp}, while those for multi-component systems are in Table \ref{tab:multicomp}.

Elasticity calculations were performed for rare-earth disilicates in the $\alpha$, $\beta$, and $\delta$ phases (corresponding to the AFLOW prototypes A2B7C2\_aP44\_2\_4i\_14i\_4i, A7B2C2\_mC22\_12\_aij\_h\_i, and A7B2C2\_oP44\_62\_3c2d\_2c\_d \cite{curtarolo:art121, curtarolo:art145, aflow_proto3}), as well as the Type A and G structures described by Felsche \cite{Felsche_Silicate_JLessCommMet_1970}.
Not all of these compositions are stable in all of these phases; in particular, the calculations indicate that Nd$_2$Si$_2$O$_7$ is elastically unstable in the $\beta$-phase (the experimentally observed stable phases for Nd$_2$Si$_2$O$_7$ are Types A and G \cite{Felsche_Silicate_JLessCommMet_1970}), with the 3rd eigenvalue of the elastic tensor being negative.
Nevertheless, it has been shown that high-entropy ceramics can incorporate components that would not be stable in the corresponding structure \cite{curtarolo:art99, curtarolo:art140, Oses_NatureReview_2020, George_NRM_2019}.
Since high-entropy disilicates frequently form in the $\beta$-phase, it is useful to know the behavior of the components in that structure, as properties of high-entropy systems often follow the rule-of-mixtures \cite{curtarolo:art140, Ridley_ActaMater_Silicates_2020, Oses_NatureReview_2020, George_NRM_2019}.
Computational approaches enable access to the properties of compositions that would be unstable in a given phase.

\begin{table*}[t!]
  \caption{\small Young's modulus, coefficient of thermal expansion and thermal conductivity for rare-earth disilicates from DFT calculations and experiment. Types A and G are indicated as A and G for brevity.
  }
  \label{tab:singlecomp}
  {\footnotesize
    \begin{tabular}{|l|c|c|c|c|c|c|c|c|c|c|c|c|c|c|}
      \hline
     \multirow{3}{*}{Compos.}  & \multirow{3}{*}{\makecell{Str. \\ type}} & \multirow{3}{*}{\makecell{Ionic \\ radii}}  &  \multicolumn{4}{|c|}{Young's modulus (GPA)} & \multicolumn{4}{|c|}{CTE ($10^{-6}$ K$^{-1}$)} & \multicolumn{4}{|c|}{Thermal Conductivity (W/m K)} \\ \cline{4-15}
      & & &   \multicolumn{2}{|c|}{DFT}    &  \multicolumn{2}{|c|}{Experiment}  &  \multicolumn{2}{|c|}{DFT}  &  \multicolumn{2}{|c|}{Experiment} & \multicolumn{2}{|c|}{DFT} & \multicolumn{2}{|c|}{Experiment} \\ \cline{4-15}
      & & & PBE & PBE+$U$ & this work &  lit.  & PBE & PBE+$U$ & this work &  lit.  & PBE & PBE+$U$ & this work &  lit.  \\ \hline
          Sc$_2$Si$_2$O$_7$ & $\beta$ & 0.745 & 187 & 224 & $229\pm14$ & 200\cite{Turcer_JEurCerS_Silicates_2018} & 28.54 & 27.36 & $5.4\pm0.7$ & 5.4\cite{Fernandez_JAmCerS_Silicates_2013} & 11.049 & 13.704 & $9.3\pm0.5$ & 9.4\cite{Turcer_ScrMater_Silicate_2018} \\
      & & & & & & 202\cite{Turcer_ScrMater_Silicate_2018} & & & & & & & & \\
     Sc$_2$Si$_2$O$_7$ & $\alpha$ & 0.83875 & 148 & & &  & 27.58 & & & & 5.904 & & & \\
     Y$_2$Si$_2$O$_7$ & $\beta$ & 0.9 & 198 & & $188\pm12$ & 170\cite{Tian_JEurCerS_Silicates_2016} & 29.54 & & $5.2\pm0.8$ & 4.1\cite{Dolan_PowdDiffr_Silicates_2008} & 11.931 & & $6.3\pm0.5$ & 5.2\cite{Turcer_ScrMater_Silicate_2018} \\
     & & & & & & & & & & & & & & 5.4\cite{Tian_JEurCerS_Silicates_2016} \\
     Y$_2$Si$_2$O$_7$ & $\alpha$ & 0.98925 & 206 & & & & 30.05 & & & & 4.862 & & & \\
     Y$_2$Si$_2$O$_7$ & $\delta$ & 0.96 & 188 & & & & 31.15 & & & & 4.052 & & & \\
     Y$_2$Si$_2$O$_7$ & G & 1.019 & 196 & & & & 30.17 & & & & 6.483 & & & \\
     La$_2$Si$_2$O$_7$ & $\beta$ & 1.032 & 109 & & & & 34.05 & & & & 4.609 & & & \\
     La$_2$Si$_2$O$_7$ & G & 1.16 & 95 & 143 & & & 34.16 & 33.32 & & 6.4 \cite{Fernandez_JAmCerS_Silicates_2013} & 2.584 & 3.354 & & \\
     La$_2$Si$_2$O$_7$ & A & 1.16 & 109 & 127 & & & 33.21 & 33.64 & & 14.0 \cite{Fernandez_JAmCerS_Silicates_2013} & 2.248 & 2.001 & & \\
     Pr$_2$Si$_2$O$_7$ & G & 1.126 & & & & &  &  & & 6.8 \cite{Fernandez_JAmCerS_Silicates_2013} &  &  & & \\
     Pr$_2$Si$_2$O$_7$ & A & 1.126 & & & & &  &  & & 11.8 \cite{Fernandez_JAmCerS_Silicates_2013} &  &  & & \\
     Nd$_2$Si$_2$O$_7$ & $\beta$ & 0.983 & 68 & & & & 30.07 & & & & 8.406 & & & \\
     Nd$_2$Si$_2$O$_7$ & A & 1.109 &  &  & $153\pm10$ & 162\cite{curtarolo:art100, APL_Mater_Jain2013} & & & $12.5\pm0.5$ & 10.5\cite{Fernandez_JAmCerS_Silicates_2013} & & & $1.2\pm0.2$ & \\
     Nd$_2$Si$_2$O$_7$ & G & 1.109 &  &  & & & & & & 6.6\cite{Fernandez_JAmCerS_Silicates_2013} & & & & \\
     Eu$_2$Si$_2$O$_7$ & $\beta$ & 0.947 & 117 & & & & 40.15 & & & & 4.948 & & & \\
     Eu$_2$Si$_2$O$_7$ & $\delta$ & 1.01 & & 119 & & & & 46.29 & & & & 1.084 & & \\
     Gd$_2$Si$_2$O$_7$ & $\beta$ & 0.938 & 143 & & & & 33.04 & & & & 6.707 & & & \\
     Gd$_2$Si$_2$O$_7$ & $\alpha$ & 1.02425 & 131 & 204 & & & 33.94 & 31.78 & & 8.3\cite{Fernandez_JAmCerS_Silicates_2013} & 2.587 & 4.540 & & \\
     Gd$_2$Si$_2$O$_7$ & $\delta$ & 1.0 & 145 & 183 & & & 33.13 & 32.7 & &  7.3\cite{Fernandez_JAmCerS_Silicates_2013} & 3.620 & 3.857 & & \\
     Gd$_2$Si$_2$O$_7$ & G & 1.053 & 113 & 179 & & & 33.37 & 31.53 & & & 3.351 & 5.121 & & \\
     Tb$_2$Si$_2$O$_7$ & $\beta$ & 0.923 & 152 & & & & 32.47 & & & & 7.063 & & & \\
     Tb$_2$Si$_2$O$_7$ & $\alpha$ & 1.01075 & 132 & & & & 33.44 & & & &  2.697 & & & \\
     Tb$_2$Si$_2$O$_7$ & $\delta$ & 0.98 & 159 & & & & 32.32 & & & & 4.140 & & & \\
     Dy$_2$Si$_2$O$_7$ & $\beta$ & 0.912 & 160 & & & & 32.29 & & & & 7.353 & & & \\
     Dy$_2$Si$_2$O$_7$ & $\alpha$ & 0.99825 & 139 & 210 &  & & 33.21 & 31.24 & &  8.5 \cite{Fernandez_JAmCerS_Silicates_2013} & 2.948 & 5.025 & & \\
     Dy$_2$Si$_2$O$_7$ & $\delta$ & 0.97 & 161 &  & & & 32.23 & & &  7.7 \cite{Fernandez_JAmCerS_Silicates_2013} & 4.301 & & & \\
     Ho$_2$Si$_2$O$_7$ & $\beta$ & 0.901 & 165 & & & & 32.13 & & & & 7.635 & & & \\
     Ho$_2$Si$_2$O$_7$ & $\alpha$ & 0.9865 & 212 & & & & 31.83 & & & & 4.445 & & & \\
     Ho$_2$Si$_2$O$_7$ & $\delta$ & 0.958 & 188 & & & & 31.67 & & & & 4.949 & & & \\
     Ho$_2$Si$_2$O$_7$ & $\gamma$ & 0.901 & & & & & & & & 4.2\cite{Fernandez_JAmCerS_Silicates_2013} & & & & \\
     Er$_2$Si$_2$O$_7$ & $\beta$ & 0.89 & 197 & & $182\pm6$ & 184\cite{APL_Mater_Jain2013,Gaillac_JPCM_Elasticity_2016} & 31.63 & & $4.3\pm0.8$ & 3.9\cite{Fernandez_JAmCerS_Silicates_2013} & 8.077 & & $4.6\pm0.4$ & \\
     Er$_2$Si$_2$O$_7$ & $\alpha$ & 0.9755 & 158 & & & & 32.68 & & & & 3.425 & & & \\
     Er$_2$Si$_2$O$_7$ & $\gamma$ & 0.89 &  & & $144\pm19$ & & & & & & & & & \\
     Er$_2$Si$_2$O$_7$ & $\delta$ & 0.945 & 169 & & & 190\cite{curtarolo:art100, APL_Mater_Jain2013} & 32.18 & & & & 4.326 & & & \\
     Yb$_2$Si$_2$O$_7$ & $\beta$ & 0.868 & 142 & 198 & $195\pm11$ & 205\cite{Turcer_JEurCerS_Silicates_2018} & 39.10 & 45.96 & $4.9\pm0.4$ & 4.0\cite{Fernandez_JAmCerS_Silicates_2013}  & 5.088 & 1.532 & $5.5\pm0.5$ & 4.3\cite{Turcer_ScrMater_Silicate_2018} \\
                                   & & & & & & 162\cite{Tian_JEurCerS_Silicates_2016} & & &  &  &  &  & & 4.45\cite{Tian_JEurCerS_Silicates_2016}  \\
     Lu$_2$Si$_2$O$_7$ & $\beta$ & 0.861 & & & $198\pm10$ & 178\cite{Tian_JEurCerS_Silicates_2016} & & & $5.4\pm0.8$ & 4.2\cite{Fernandez_JAmCerS_Silicates_2013} &  & & $8.7\pm0.7$ & 4.3\cite{Turcer_ScrMater_Silicate_2018} \\
                                   &&  & & & & & & & & &  & & & 4.4\cite{Tian_JEurCerS_Silicates_2016} \\
      \hline
    \end{tabular}
  }
\end{table*}

\begin{table*}[t!]
  \caption{\small Young's modulus, coefficient of thermal expansion and thermal conductivity for multi-component rare-earth disilicates from experiment.
  }
  \label{tab:multicomp}
  {\footnotesize
    \begin{tabular}{|l|c|c|c|c|c|c|c|c|}
      \hline
     \multirow{2}{*}{Composition}   &  \multirow{2}{*}{\makecell{Structure \\ type}} & \multirow{2}{*}{\makecell{Ionic \\ radii}}   & \multicolumn{2}{|c|}{Young's modulus (GPA)}& \multicolumn{2}{|c|}{CTE ($10^{-6}$ K$^{-1}$)} & \multicolumn{2}{|c|}{Thermal Conductivity (W/m K)} \\ \cline{4-9}
      & & & Measured &  ROM & Measured & ROM & Measured & ROM \\ \hline
     (Sc$_{0.33}$Yb$_{0.67}$)$_2$Si$_2$O$_7$ & $\beta$ & 0.827 & $206\pm10$ & 206 & $4.7\pm0.6$ & 5.1 & $3\pm0.3$ & 6.8 \\
      (ScYYb)$_2$Si$_2$O$_7$ & $\beta$ & 0.838 & $196\pm15$ & 204 & $4.5\pm0.6$ & 5.2 & $2.2\pm0.3$ & 7 \\
      (ErYbLu)$_2$Si$_2$O$_7$ & $\beta$ & 0.873 & $191\pm15$ & 192 & $5.6\pm0.7$ & 4.9 & $6.3\pm0.5$ & 6.3 \\
      (YErYbLu)$_2$Si$_2$O$_7$ & $\beta$ & 0.880 & $185\pm15$ & 191 & $4.9\pm0.6$ & 4.95 & $4.3\pm0.4$ & 6.3 \\
      (Sc$_{0.15}$Y$_{0.4}$Er$_{0.5}$Yb$_{0.5}$Lu$_{0.45}$)Si$_2$O$_7$ & $\beta$ & 0.869 & $204\pm17$ & 194 & $4.1\pm0.4$ & 4.96 & $2.5\pm0.3$ & 6.4 \\
      (ScNdErYbLu)$_2$Si$_2$O$_7$ & Multi-phase & 0.869 & $179\pm16$ & & $4.8\pm0.3$ & & &  \\
      (YYb)$_2$Si$_2$O$_2$ & $\beta$ & 0.884  &    &    & 3.5 \cite{Wang_Silicates_JAmCerS_2022} &    &  4.2 \cite{Wang_Silicates_JAmCerS_2022} &   \\
       &  &   &    &    &  &    &  3.6 \cite{Turcer_Silicates_ScrMater_2021} &   \\
      (Y$_{0.1}$Yb$_{0.9}$)$_2$Si$_2$O$_2$ & $\beta$ & 0.8712 &    &   &   &    & 4.2 \cite{Turcer_Silicates_ScrMater_2021} &   \\
      (YYbEr)$_2$Si$_2$O$_2$ & $\beta$ & 0.886  &    &    & 3.3 \cite{Wang_Silicates_JAmCerS_2022}  &    & 4.2 \cite{Wang_Silicates_JAmCerS_2022}  &   \\
      (YYbErSc)$_2$Si$_2$O$_2$ & $\beta$ & 0.85075 &    &    & 3.2 \cite{Wang_Silicates_JAmCerS_2022}  &    & 3.9 \cite{Wang_Silicates_JAmCerS_2022} &   \\
      (YYbErScGd)$_2$Si$_2$O$_2$ & $\beta$ & 0.8682 &    &    & 3.2 \cite{Wang_Silicates_JAmCerS_2022} &    & 3.1 \cite{Wang_Silicates_JAmCerS_2022} &   \\
      (YYbErScGdEu)$_2$Si$_2$O$_2$ & $\beta$ & 0.8133 &    &    & 3 \cite{Wang_Silicates_JAmCerS_2022} &    & 2.3 \cite{Wang_Silicates_JAmCerS_2022} &   \\
      (YYbScGdLu)$_2$Si$_2$O$_2$ & $\beta$ & 0.867 &    &    &    &    & 2.2 \cite{Wang_Silicates_JAmCerS_2022} &   \\
      \hline
    \end{tabular}
  }
\end{table*}

The elastic moduli are expected to decrease with increasing $RE$ ionic radius: the rare-earth metal-oxide bond strength decreases with increasing ionic radius for a given ionic charge state and coordination number, and the dependence of elastic properties on volume per atom is well-known \cite{curtarolo:art124}.
This appears to be the case here for both the calculations and the experiment: the elastic moduli generally decrease with increasing radii of the rare-earth cations, with Pearson correlation values of -0.47 for the calculations (lowest magnitude for the $\alpha$-phase at -0.002, with larger values of -0.69 and -0.75 for the $\beta$ and $\delta$ phases).
If only the preferred stable ground states are considered, the correlation is -0.6 (-0.63 and -0.14 for the $\alpha$ and $\beta$ phases), while for the unstable phases the value is -0.41 (-0.79 and -0.75 for the $\beta$ and $\delta$ phases).
The particularly large deviation from the trend for Nd$_2$Si$_2$O$_7$ could be explained by its elastic instability in the $\beta$-phase, as discussed above, resulting in a relatively low elastic modulus of 68 GPa.
Elastic moduli for the other materials range from 95 GPa for La$_2$Si$_2$O$_7$ in the Type G structure up to 212 GPa for $\alpha$-phase Ho$_2$Si$_2$O$_7$; for other $\beta$-phase materials, elastic moduli range from 109 GPa for La$_2$Si$_2$O$_7$ to 198 GPa for Y$_2$Si$_2$O$_7$.
The linear fit between the elastic moduli $E$ and the ionic radii $r_{\mathrm{i}}$  for all systems calculated with PBE is given by the equation $E = 346 - 199 r_{\mathrm{i}}$, with a relative RMS deviation in the fit of 0.13.

For the experimental results, a large Pearson correlation of -0.749 is observed between $E$ and $r_{\mathrm i}$ for the single-component materials, increasing to -0.982 when just the $\beta$ phase results are considered.
Nd$_2$Si$_2$O$_7$ is again an outlier: it has a significantly larger cation radius than any of the other materials, and is the only stable material here with the Type A structure \cite{Felsche_Silicate_JLessCommMet_1970}.
The correlation for the multi-component materials is -0.621, while the correlations for all of the experimental elastic moduli is -0.747.
The linear fit between the experimentally measured elastic moduli $E$ and the $RE$ ionic radii $r_{\mathrm{i}}$ for all single-component systems is given by the equation $E = 362 - 198 r_{\mathrm{i}}$, with a relative RMS deviation in the fit of 0.07.

The elastic moduli of the multi-component systems are generally similar to what would be expected from the rule of mixtures, as listed in Table \ref{tab:multicomp}.
In Figure \ref{fig:elasticmod}(b), the values for the multi-component systems are in the same region of the plot as the values for the single-component $\beta$ phases.
The correlation between the measured values and those predicted from the rule of mixtures is 0.629, while the RMSrD is 0.036.

\subsection*{Thermal Expansion}

\begin{figure*}[t!]
  \includegraphics[width=0.98\textwidth]{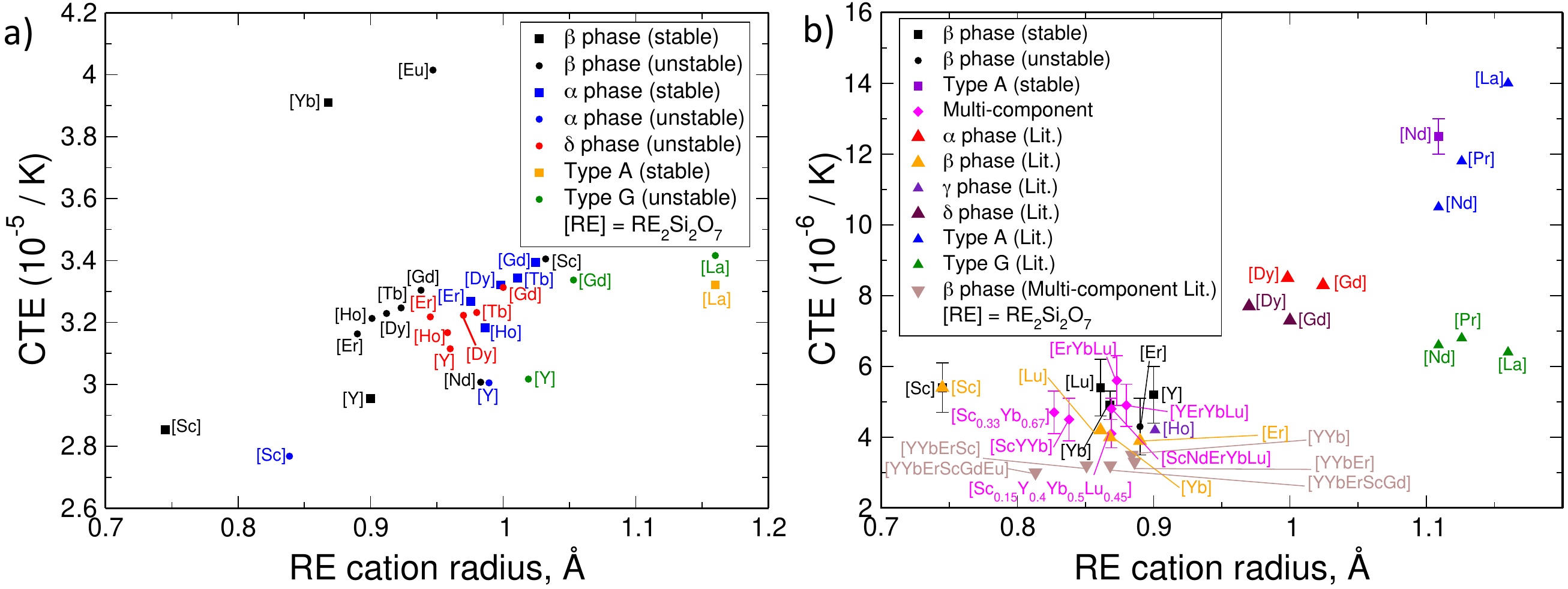}
  \vspace{-4mm}
  \caption{\small Coefficient of thermal expansion as a function of rare-earth ($RE$) cation radius for rare-earth disilicates obtained from
    {\bf (a)}  DFT calculations and
    {\bf (b)} experimental measurements, including data extracted from the literature \cite{Fernandez_JAmCerS_Silicates_2013, Wang_Silicates_JAmCerS_2022}.}
  \label{fig:thermalexpansion}
\end{figure*}

\noindent
Silicate materials display varied CTE, despite containing rigid Si-O bonds with a thermal expansion coefficient near zero \cite{Cameron_AmMin_Silicate_1973}.
It can be assumed that the Si-O tetrahedra do not contribute significantly to thermal expansion in rare-earth disilicate materials and that the rare-earth metal-oxygen bonding should govern thermal expansion behavior \cite{Hazen_AmMin_Silicate_1977}.
The $RE$ metal-oxide bond strength decreases with increasing ionic radius for a given charged ion and coordination number, leading to the thermal expansion of disilicates increasing with the $RE$ cation radius.
This trend can be observed in the CTE results from both DFT calculations and experimental measurements, along with CTE data extracted from the literature from dilatometry and high-temperature XRD techniques for $RE_2$Si$_2$O$_7$ \cite{Fernandez_JAmCerS_Silicates_2013, Wang_Silicates_JAmCerS_2022, Strzelecki_ACSESC_Silicates_2020, Ayyasamy_JAmCerS_Silicates_2020}, listed in Tables \ref{tab:singlecomp} and \ref{tab:multicomp}, and plotted against the average $RE$ cation radius in Figure \ref{fig:thermalexpansion}.

For the calculated values, the Pearson correlation between the CTE and $RE$ ionic radii is 0.26 (lowest magnitude for the $\beta$-phase at 0.27, with higher values of 0.88 and 0.71 for the $\alpha$ and $\delta$ phases).
If only the preferred stable ground states are considered, the correlation is 0.29 (0.82 and 0.4 for the $\alpha$ and $\beta$ phases), while for the unstable phases the value is 0.29 (0.1 and 0.71 for the $\beta$ and $\delta$ phases).
The linear fit between the CTE for all systems calculated with PBE and the $RE$ ionic radii $r_{\mathrm{i}}$ is given by the equation $\mathrm{CTE} = 2.472 + 0.803 r_{\mathrm{i}}$, with a relative RMS deviation in the fit of 0.081.

For the experimental values for the single-component systems measured in this work, the Pearson correlation between the CTE and $RE$ ionic radii is 0.843, although this might be skewed by the large difference between the Nd$_2$Si$_2$O$_7$ Type A value and the results for the $\beta$ phase materials; for just the $\beta$ phase systems, the correlation is -0.504.
The correlation for the multi-component materials measured in this work is 0.298, while the correlations for all of the experimental CTE is 0.969.
When the values from Ref. \onlinecite{Fernandez_JAmCerS_Silicates_2013} are included, the correlation between the CTE and $RE$ ionic radii is 0.821, dropping to 0.73 when just single-component materials are considered.
The correlation for just the results from Ref. \onlinecite{Fernandez_JAmCerS_Silicates_2013} is 0.697, with correlations of 0.999 for Type A, -0.996 for the $\beta$ phase, and -0.655 for Type G.
The correlation of the CTE with the ionic radius for the multi-component materials from Ref. \onlinecite{Wang_Silicates_JAmCerS_2022} is 0.876; when included with the rest of the experimental results the correlation is 0.777.
The linear fit between the experimentally measured CTE and the $RE$ ionic radii $r_{\mathrm{i}}$ for all single-component systems is given by the equation $\mathrm{CTE} = -13.3 + 21.85 r_{\mathrm{i}}$, with a relative RMS deviation in the fit of 0.068; changing to $\mathrm{CTE} = -8.9 + 16.36 r_{\mathrm{i}}$ with an RMS deviation of 0.1 when the results from Ref. \onlinecite{Fernandez_JAmCerS_Silicates_2013} are included.

Similar to the elastic moduli, the CTE of the multi-component systems are generally similar to what would be expected from the rule of mixtures, as listed in Table \ref{tab:multicomp}.
In Figure \ref{fig:thermalexpansion}(b), the values for the multi-component systems are in the same region of the plot as the values for the single-component $\beta$ phases.
The correlation between the measured values and those predicted from the rule of mixtures is -0.445, while the RMSrD is 0.143.

\subsection*{Thermal Conductivity}

\begin{figure*}[t!]
  \includegraphics[width=0.98\textwidth]{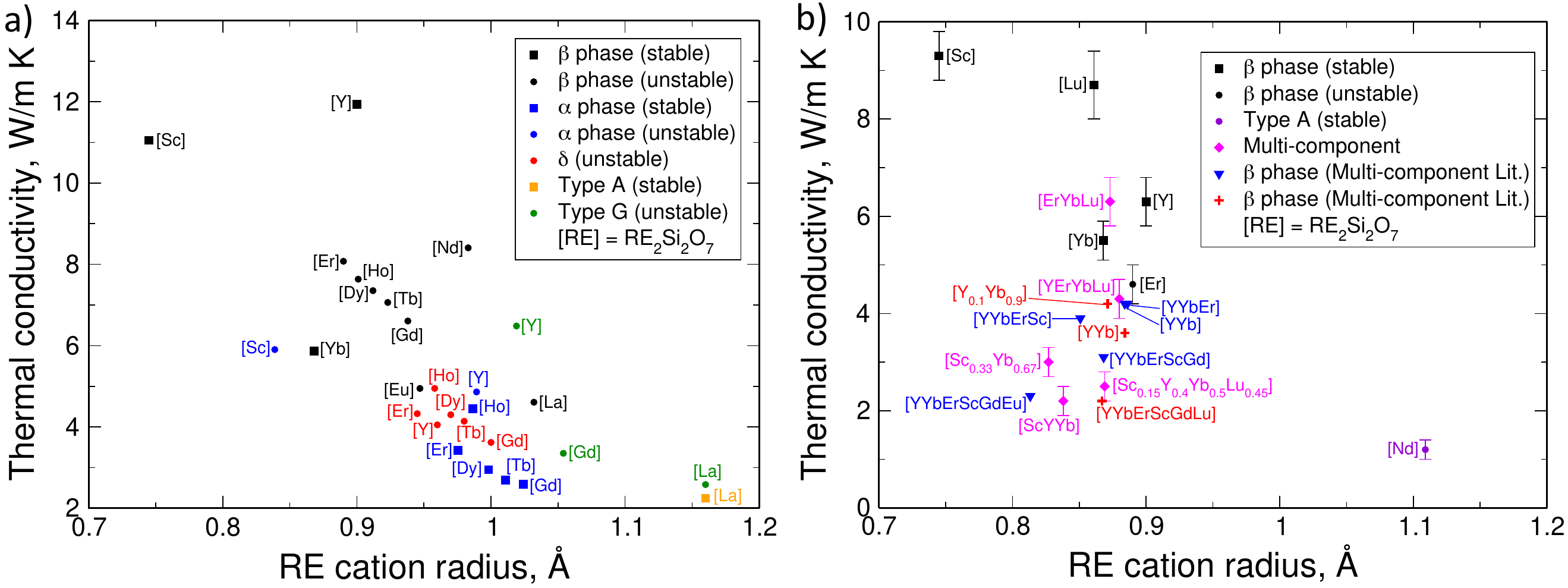}
  \vspace{-4mm}
  \caption{\small Room temperature thermal conductivity as a function of rare-earth ($RE$) cation radius for rare-earth disilicates obtained from
    {\bf (a)}  DFT calculations and
    {\bf (b)} experimental measurements, including data extracted from the literature \cite{Wang_Silicates_JAmCerS_2022, Turcer_Silicates_ScrMater_2021}.}
  \label{fig:thermalcond}
\end{figure*}

\noindent
Figure \ref{fig:thermalcond} shows the thermal conductivity for rare-earth disilicates from DFT calculations and SSTR measurements at room temperature.
For the calculated values, the Pearson correlation between the thermal conductivity and $RE$ ionic radii is -0.72 (lowest magnitude for the $\beta$-phase at -0.6, with higher values of -0.82 and -0.68 for the $\alpha$ and $\delta$ phases).
If only the preferred stable ground states are considered, the correlation is -0.8 (-0.72 and -0.19 for the $\alpha$ and $\beta$ phases), while for the unstable phases the value is -0.63 (-0.52 and -0.68 for the $\beta$ and $\delta$ phases).
The linear fit between the thermal conductivity $\kappa$ for all systems calculated with PBE and the $RE$ ionic radii $r_{\mathrm{i}}$ is given by the equation $\kappa = 25.66 - 20.98 r_{\mathrm{i}}$, with a relative RMS deviation in the fit of 0.068.

The correlation between the experimental thermal conductivity and $RE$ ionic radii for the single-component materials is -0.911, although once again this value might be skewed by the very different results for Type A Nd$_2$Si$_2$O$_7$ compared to the $\beta$ phase; for just the $\beta$ phase systems, the correlation is -0.746.
The correlation for the multi-component materials measured in this work is 0.576, while the correlations for all of the thermal conductivities experimentally measured in this work is -0.552.
For the results in Ref. \onlinecite{Wang_Silicates_JAmCerS_2022} the correlation is 0.849; while the correlation for the results in Ref. \onlinecite{Turcer_Silicates_ScrMater_2021} is 0.45. When these data are included with the multicomponent materials measured in this work, the correlation is 0.53, while the correlation for all of the materials is -0.42.
The linear fit between the experimentally measured thermal conductivity $\kappa$ for single-component materials and the $RE$ ionic radii $r_{\mathrm{i}}$ is given by the equation $\kappa = 26.2 - 22.67 r_{\mathrm{i}}$, with a relative RMS deviation in the fit of 0.043.

\begin{figure}[t!]
  \includegraphics[width=0.48\textwidth]{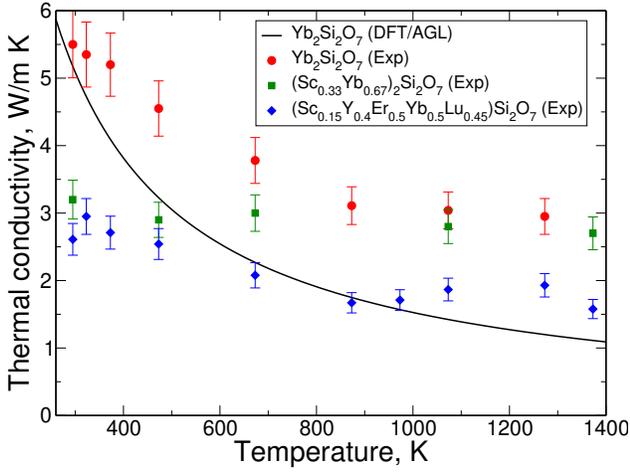}
  \vspace{-4mm}
  \caption{\small Thermal conductivity as a function of temperature for rare-earth disilicates obtained from
 DFT calculations and experimental measurements.}
  \label{fig:thermalcondht}
\end{figure}

Thermal conductivity as a function of temperature up to 1100$^\circ$C are shown for (Sc$_{0.15}$Y$_{0.4}$Er$_{0.5}$Yb$_{0.5}$Lu$_{0.45}$)Si$_2$O$_7$, (Sc$_{0.33}$Yb$_{0.67}$)$_2$Si$_2$O$_7$ and Yb$_2$Si$_2$O$_7$ in Figure \ref{fig:thermalcondht}. For the computational results for Yb$_2$Si$_2$O$_7$, the temperature dependence is obtained using the $\kappa(T) \sim 1/T$ approximation \cite{slack, Morelli_Slack_2006,curtarolo:art96, curtarolo:art115}. It is notable that while the experimentally-obtained thermal conductivity for the single-component system reduces quite significantly with increasing temperature, the results for the multi-component systems are less affected by temperature.

The strong dependence of the thermal conductivity on the $RE$ ionic radii is to be expected for the following reasons:
(i) as discussed above, bond strengths and thus elastic moduli in the material depend strongly on the $RE$ metal-oxygen bond length, which in turn affects phonon propagation speeds;
(ii) the ionic radius depends strongly on the cation coordination numbers, which in turn are known to correlate with the anharmonicity of the material \cite{Miller_Anharmonicity_ChemMater_2017}.
Comparisons between the thermal conductivity and the $RE$ cation mass were also performed, but the trend was not as strong as for the ionic radii values, with correlations of -0.426 for the calculations and -0.379 for the experiments.

Note that in comparison to the elastic moduli and CTE, the thermal conductivity of the multi-component systems are generally significantly lower than what would be predicted by the rule of mixtures, as listed in Table \ref{tab:multicomp}: the correlation between the values measured in this work and those predicted by the rule of mixtures is -0.68, while the RMSrD is high (indicating large differences) at 0.806.
This result is in agreement with observations for other high entropy materials \cite{Braun_ESO_AdvMat_2018}, and is likely due to the configurational disorder in the interatomic force constants leading to increased phonon scattering and lower thermal conductivity \cite{Braun_ESO_AdvMat_2018}.

\section*{Discussion}

\noindent
The trends demonstrated in this work for thermal and elastic properties with respect to ionic radii from both DFT calculations and experimental measurements facilitate the prediction of material properties for typically unstable polymorphs, where the properties cannot be directly measured or reliably calculated.
Design rules based on these trends will be useful for optimizing environmental barrier coatings through stabilization of rare-earth elements into typically unstable $RE_2$Si$_2$O$_7$ polymorphs, such as $RE$ = La-Er stabilization into the low CTE $\beta$ polymorph over the entire temperature range of interest for turbine hot section components.
While phase stabilization has been demonstrated in the literature \cite{Fernandez_JSSChem_Silicates_2011, Dong_JECerS_2019, Sun_CorSci_HESilicates_2020}, the impact of multi-component $RE_2$Si$_2$O$_7$ solutions on all pertinent EBC material properties is not yet fully understood.
Property trends across all rare-earth elements should be analyzed to help predict properties when stabilizing thermodynamically unstable phases.

While the elastic moduli and CTE for multi-component systems generally follow the rule of mixtures, the thermal conductivity is generally significantly lower than what would be predicted, in accordance with previous observations for oxides \cite{Braun_ESO_AdvMat_2018}.
Therefore, the multi-component high-entropy approach enables the design of rare-earth disilicate ETBCs that combine high stiffness and low thermal conductivity, with a CTE engineered to match that of the SiC substrate.

\section*{Conclusion}

\noindent
A combination of first-principles calculations and experimental measurements were used to identify trends in material properties for the rare-earth disilicate systems.
The results show that the Young's modulus, coefficient of thermal expansion, and thermal conductivity all correlate with $RE$ cation
radius for $RE_2$Si$_2$O$_7$ materials.
The search for next-generation environmental barrier coating candidates should thus focus on rare-earth disilicates with smaller $RE$ ionic radii, or a higher Pauling bond strength, as a method for maintaining a low coefficient of thermal expansion.
From the viewpoint of crystal structure, stabilization of $RE_2$Si$_2$O$_7$ into the $\beta$ phase should also decrease the $RE$---O coordination number, where a lower coordination number also correlates with a decreased coefficient of thermal expansion value.
Based on the observed trends, multi-component disilicates represent a promising avenue of future research for tailoring low thermal expansion environmental barrier coatings.

\noindent{\small\textbf{Acknowledgments.}}
Research sponsored by NSF (DMR-2219788, DMR-1921973 and DMR-1921909) and Office of Naval Research (N00014-21-1-2477).

\noindent{\small\textbf{Author contributions.}
The authors contributed equally to the article.
}

\noindent{\small\textbf{Competing interests.} The authors declare no competing interests.}

\newcommand{\Ozolins}{Ozoli{\c{n}}{\v{s}}}

\end{document}